\documentclass[aps,twocolumn,prl,showpacs,superscriptaddress,groupedaddress]
{revtex4-1}  
\usepackage{graphicx}
\usepackage{dcolumn}  
\usepackage{bm}  
\usepackage{amssymb}
\usepackage{amsmath}

\makeatother

\newcommand{\nnabla}{\mathbf \nabla}

\newcommand{\req}[1]{Eq.~(\ref{#1})}

\newcommand{\rref}[1]{(\ref{#1})}

\renewcommand{\r}{\mathbf{r}}
\renewcommand{\k}{\mathbf{k}}

\renewcommand{\a}{\mathbf{a}}\renewcommand{\v}{\mathbf{v}}
\newcommand{\pt}{\partial_t}

\newcommand{\beq}{\begin{equation}}
\newcommand{\eeq}{\end{equation}}
\newcommand{\be}{\begin{equation}}
\newcommand{\ee}{\end{equation}}
\newcommand{\beqa}{\begin{eqnarray}}
\newcommand{\eeqa}{\end{eqnarray}}
\newcommand{\bea}{\begin{eqnarray}}
\newcommand{\eea}{\end{eqnarray}}
\usepackage{amssymb}
\usepackage{array}
\usepackage{amsmath}
\usepackage{graphicx}\usepackage{graphics}
\usepackage{dcolumn}
\usepackage{bm}\usepackage{varioref}

\renewcommand{\b}{{\mathbf b}}\renewcommand{\t}{{\mathbf t}}
\newcommand{\cE}{{\cal E}}
\newcommand{\slabel}[1]{\tag{S.\theequation}\label{#1}
\addtocounter{equation}{1}}

\begin{document}
\newcommand{\niceref}[1] {Eq.~(\ref{#1})}
\renewcommand{\fullref}[1] {Equation~(\ref{#1})}
\title{Turning the BCS-BEC {\em crossover} into a phase {\em transition} by radiation. }

\date{\today}

\begin{abstract}

We show that the  Bardeen-Cooper-Schrieffer state (BCS)   and the Bose-Einstein condensation (BEC)
sides of the BCS-BEC crossover can be rigorously distinguished from each other 
by the extrema of the spectrum of the fermionic excitations. Moreover, we demonstrate that this formal distinction is realized  
as a non-equilibrium phase transition under  radio frequency radiation. The BEC phase
remains translationally invariant, whereas the BCS phase transforms  into the supersolid phase.
For a two-dimensional system this effect occurs at arbitrary small amplitude of the radiation field. 

% The Bose-Einstein condensation (BEC) to Bardeen-Cooper-Schriffer state (BCS)  transition in 
% cold Fermi gases is a paradigm of a thermodynamic crossover. Two different states may be adiabatically connected without any thermodynamic singularity.  
% With the recent development of the controllable Feshbach resonance in cold atomic gases this crossover may be now controllably explored.  We show that by applying a weak radio frequency perturbation in a two-dimensional cold atomic gas, this crossover may transmuted into a non-equilibrium phase transition 
% between  the usual superfluid state and a supersolid state that spontaneously breaks translational symmetry. The transition would provide an interesting arena to study 
% non-equilibrium phase transitions, the kinetics of first-order quantum phase transitions, the properties of the supersolid 
% as well as illuminating the properties of the underlying BEC-BCS crossover.

\end{abstract}

\pacs{67.85.-d;
67.85.Lm; 
67.80.-s;
03.75.Kk 	 }
\author{Yonah Lemonik}
\email{lemonik@phys.columbia.edu}

\author{Igor L. Aleiner}
\email{aleiner@phys.columbia.edu}

\author{Boris L. Altshuler}
\email{bla@phys.columbia.edu}

\affiliation{Department of Physics, Columbia University, New York, NY, 10027, USA}
\maketitle
{\em Introduction}-- 
The Bose-Einstein condensation (BEC) \cite{BECReview} and Bardeen-Cooper-Schrieffer state (BCS) \cite{BCS} are two extreme scenarios
for the formation of the superfluid state in fermionic systems where 
the only allowed gapless mode is the acoustic 
bosonic branch.
 In the BEC scenario,
the fermions are first paired into compact two-particle complexes (molecules). These molecules
experience Bose-Einstein condensation with the acoustic low energy spectrum due to the weak repulsion
between molecules. In the BCS scenario, weakly coupled Cooper pairs are formed from states
near the Fermi level so that the characteristic size for the pair correlation significantly exceeds  
the interparticle distance. Nevertheless, this weak coupling is sufficient to gap
the fermionic excitation and leads to bosonic acoustic excitations as the oscillations of the
order parameter. The physical effects occurring in between those two scenarios are referred to as BCS-BEC {\em crossover}. 

The BCS-BES crossover is captured by the simplest 
Hamiltonian density \cite{BECBCSTheory} 
\begin{subequations}
\label{eq:Hferm}
\be
H=\psi^*_\sigma h^{(1)}\psi_\sigma + b^*h^{(2)}b +
\left[\frac{\lambda}{2} b^*\psi_{\sigma_1} \tau^y_{\sigma_1\sigma_2}\psi_{\sigma_2} +{\mathrm h.c.}\right],
\label{eq:Hferm-a}
\ee
where $\sigma$ labels two (spin or pseudospin) states for the fermions described
by Grassmann fields $\psi_\sigma(\r), \psi_\sigma^*(\r)$
(summation over repeated indices is implied, and $\tau^{y}$ is the standard Pauli matrix), 
and the bosonic fields $b(\r),b^*(\r)$ describe the bound states of two fermions.
The single specie energy part is described by ($\hbar=1$)
\be 
h^{(q)}\equiv
\left(-i\nnabla-q\a\right)^2/({2qm})+q\varphi - \delta_{2,\,q}\varepsilon_b,
\label{eq:Hferm-b}
\ee
where the background vector $\a(\r,t)$ and scalar potentials $\varphi(\r,t)$ are introduced to highlight
the continuity equation for the total particle density  
$n(\r)=  \psi_\sigma^* \psi_\sigma+2b^*b$.
\end{subequations}

The parameter $\varepsilon_b$ describes the energy of the bound state when $\varepsilon_b>0$ or the position of the
resonance when $\varepsilon_b<0$. 
In models of pre-formed pairs in superconductors \cite{PreformedPairs}, $\varepsilon_b$
is a material dependent parameter. In experiments with cold atoms $\varepsilon_b$ is the directly tunable position of the Feshbach resonance
 \cite{FeshbachReview2010,ColdAtomReview}. Thus, cold atom system provide
a versatile platform for a detailed study of the BEC-BCS crossover.
 
In \req{eq:Hferm-b}, the constant $\lambda$ controls the coupling of the bound state (molecules) with fermionic continuum.
If  $\lambda$ is sufficiently small $n^{2/D}/m \gg |\lambda| n^{1/2}$ (so-called narrow resonance regime), an analytic
treatment of the problem is possible for any $\epsilon_b$, and dimensionality $D$ \cite{Supplemental}. For large $\lambda$
the crossover can be investigated only numerically \cite{2DMonteCarlo}.

{\em Definition of the ``critical field'' of the crossover} --
The arguments below are independent of the width of the resonance. 
All numerical and analytical study of the ground state energy of the
Hamiltonian \rref{eq:Hferm-a} at $\varphi,\a=0$ indicates that the ground state energy density $E_{GS}(n,\varepsilon_b)$ is an analytic function of its arguments, (hence the term {\em crossover} rather than {\em transition}). The usual argument is that regardless of the values of the parameters, the last term
in \req{eq:Hferm-a} leads to an anomalous average $\langle \psi_{\sigma_1}\tau^y_{\sigma_1\sigma_2}\psi_{\sigma_2}\rangle\propto \langle b\rangle \propto e^{i2\theta}$. 
Given that apparently no other symmetry breakings occur, there is no sharply defined critical  field $\varepsilon_b^c(n)$ that separates the BEC and BCS regimes.
Moreover,  for all parameters the low energy excitations are described by superfluid hydrodynamics given by the Lagrangian
\be
{\cal L}=\!
n\left[\phi - \frac{\v^2}{2m}\right]\! -  E_{GS}(n)
+\gamma^*_\sigma\left[i\tilde{\pt}-\epsilon(-i\nnabla;n)\right] \gamma_\sigma,
\label{hydrodynamics}
\ee
where $\tilde{\pt}\equiv \pt+\v\cdot\nnabla$ is the convective derivative, $\phi\equiv \partial_t \theta+\varphi$ and $\v\equiv \nnabla \theta -\a $ are the gauge invariant potential and velocities respectively,
 the fields $n$ and $\theta$ are real, and $\gamma_\sigma,\gamma_\sigma^*$ are the Grassman fields
describing the fermionic excitations (which for the problem of interest can be viewed as neutral BCS quasiparticles). 
\begin{figure}[h]
\includegraphics[width = 0.9\columnwidth]{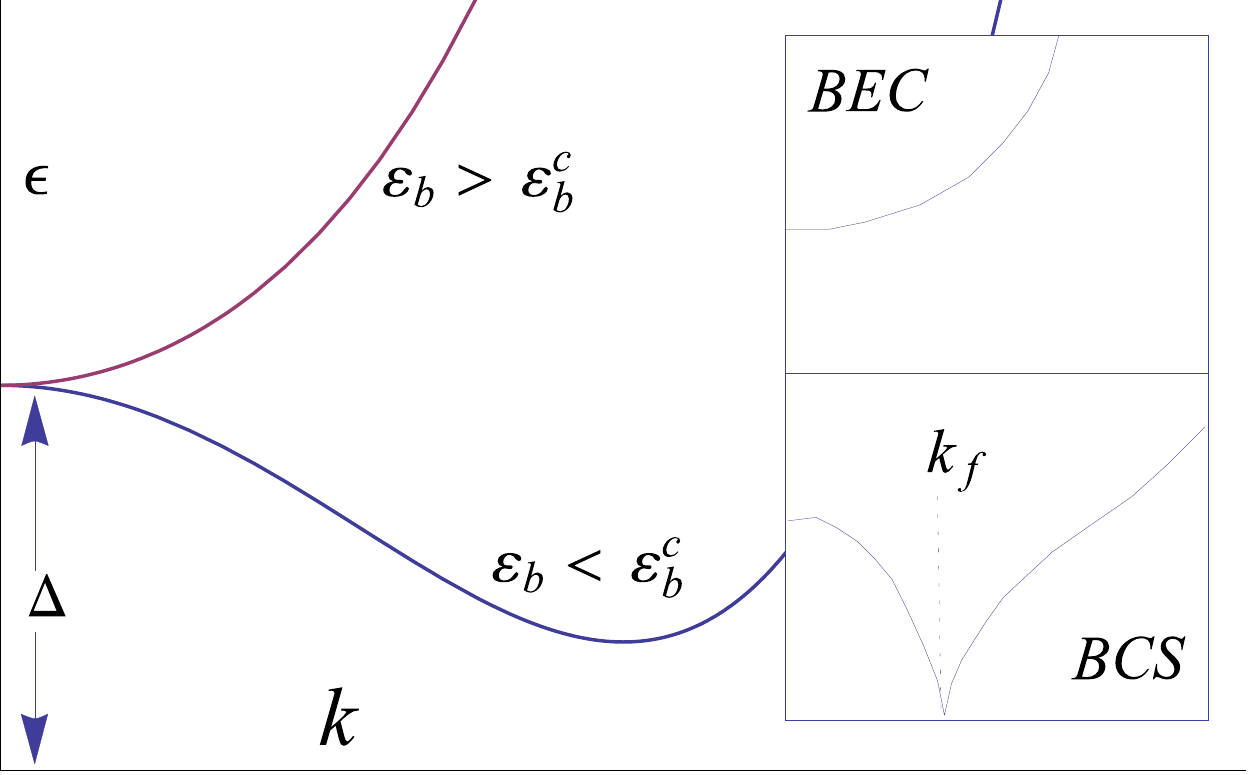}
\caption[The spectrum of fermionic quasi-particles near critical detuning]{ The fermionic quasiparticle spectrum $\epsilon(k)$ from \req{hydrodynamics} for $\varepsilon_b$ close to its critical value \rref{critical-field}.
Insets are the fermionic spectra deep in the BEC and BCS regimes.}
\label{fig:BogoSpec}
\label{fig1}
\end{figure}
This Lagrangian is an analytic function of variables, which apparently does not allow a definition of a critical field separating the two regimes.
The first term in  Lagrangian \rref{hydrodynamics} is protected by
the gauge and Galilean invariances and the second term, $ E_{GS}(n;\epsilon_b)$, is an analytic function of $\epsilon_b$. This implies that the spectrum of the bosonic excitations (phonons) also must be analytic.  However, the spectrum of the fermionic excitations $\epsilon(k)$ experiences a reconstruction that allows for the definition of the critical  field 
$\varepsilon_b^c(n)$. 

In the deep BEC regime the fermions are entirely decoupled from bosons so that the spectrum has a minimum at $k=0$.
In the opposite limit, in the deep $BCS$ regime, the spectrum of the quasiparticles has minima on the Fermi surface $k=k_F$, see inset to Fig.~\ref{fig1}. As the transition from
point to sphere can not be analytic, there must exist a point $\varepsilon_b^c(n)$ such that 
\begin{subequations}
\label{definitions}
\be
\left.{d^2\epsilon(k;\varepsilon_b,n)}/{dk^2}\right|_{k=0,\ \varepsilon_b=\varepsilon_b^c(n)}=0.
\label{critical-field}
\ee  
We will call $\varepsilon_b^c(n)$ from \req{critical-field}  the {\em critical field} of the BEC-BCS {\em transition}.  
The fermionic spectrum for small momenta can be written as
\be
\epsilon(k,n)=\Delta(n,\varepsilon_b)+{\alpha \left(\varepsilon_b-\varepsilon_b^c\right)k^2}/{2} +{ \beta k^4}/{4}
\label{spectrum-expanded}
\ee
where $\alpha,\beta>0$. At fields below the transition $\varepsilon_b<\varepsilon_b^c$, the spectrum is Mexican hat shaped with the position of the minimum
$|k|=k_F$, and its energy $\Delta-\Lambda$:
\be
k_F=\sqrt{(\alpha/\beta)\left(\varepsilon_b^c-\varepsilon_b\right)}, \quad  \Lambda=\beta k_F^4/4
\label{kF}
\ee
\end{subequations}

At first glance, the definition \rref{definitions} appears to be of no physical consequence. Indeed, at $\varepsilon_b=\varepsilon_b^c$ the fermionic spectrum remains gapped
so that there is neither a reconstruction of the ground state nor a thermodynamic singularity at finite temperature. However we now show 
that by an arbitrarily weak time-dependent perturbation  it is possible to induce a spontaneous symmetry breaking 
of the ground state of the two-dimensional system precisely at the critical point \rref{critical-field}.
For a finite perturbation, the theory outlined below predicts the formation of the incommensurate {\em supersolid} state via a weak quantum first order phase transition. 
The periodicity of this phase will be determined by the ``order parameter'' \rref{kF}. 

{\em Turning the crossover to the transition by coupling to radiation} --
The controlled radiative coupling to the ``external'' fermions has been experimentally demonstrated, for example in Ref.~\cite{RF}. It involves a third species of fermions described
by the Grassmann fields $f^*,f$ which originally do not interact with any of the particles of the original problem \rref{eq:Hferm}. In the context of cold atom systems this would be given by a third hyperfine state. The radiation induces transitions  between
the third species
and one of the fermions from \req{eq:Hferm}, so that in terms of the low energy theory \rref{hydrodynamics} it creates or annihilates two fermionic excitations, 
\be
H_{RF}=f^*\left(h^{(1)} +\Delta_f\right)f + \left[F_\sigma(t)e^{-i\theta}  f\gamma_\sigma +h.c.\right],
\label{HRF}
\ee
where $F(t)$ is proportional to the strength of the radiation field, and $\Delta_f>0$ is the boundary of
the spectrum of the third specie. The functional form of the second term in \req{HRF} is protected by gauge invariance.
Let us concentrate on the case of the monochromatic radiation $F(t) \propto e^{i\omega t}$,
and introduce detuning as
\be
d\equiv \Delta(n,\varepsilon_b)+\Delta_f-\omega.
\label{detuning}
\ee
If $d<0$ the Hamiltonian \rref{HRF} creates {\em real} fermion pairs and phonons and therefore leads to entropy
growth (heating). For $d>0$ real processes are not allowed 
(in fact, multi-photon real processes are also forbidden
as the Hamiltonian \rref{HRF} necessarily creates one $f$ particle per
one photon 
), and the coupling 
\rref{HRF} introduces a correction to the ground state energy density
\be
\delta E_{GS}^{(0)}= \int \frac{d^2 k }{(2\pi)^2}\cE(k);
\quad \cE(k)=-
\frac{F_\sigma F^*_\sigma}{\xi(k)},
\label{correction1}
\ee
where $\xi(k)\equiv {k^2/(2m)+\epsilon(k)-\Delta+ d}$ is the energy of the virtual state consisting of two excited fermions, 
and
the meaning of the superscript $(0)$ will become clear shortly.

The correction $\delta E_{GS}^{(0)}$ is logarithmically divergent as $d\to 0$. This corresponds to a photon with energy just sufficient for the excitation of  $f$ and $\gamma$ fermions with zero momentum.
Above the BCS-BEC {\em transition} field, $\varepsilon_b > \varepsilon_b^c$. This is the lowest energy of {\em any} excitation
of a $f$ and a $\gamma$ fermion and  \rref{correction1} is the final answer. As this energy correction
by itself is not observable, radiation above the threshold, $d>0$, does not lead to any changes in the 
properties of the ground states of the system.

The situation changes qualitatively below the BCS-BEC {\em transition}, $\varepsilon_b < \varepsilon_b^c$. Indeed the minimal energy
of the pair excitations is given by a fermion $f$ with $k=0$ and one of the $\gamma$ fermions at $|k| = k_F$ from \req{kF},
i.e. the lowest boundary of the two particle continuum is given by $\Delta(n)+\Delta_f-\Lambda$. If this were to appear in the denominator in \req{correction1}  this would give a divergence already at $d=\Lambda$, and a singular correction to the ground state.
For the homogeneous state this is impossible as the translational invariance of the 
ground state and of the Hamiltonian \rref{HRF} prohibits the excitation of two-quasiparticle with
the total momentum $|k|=k_F$ from a zero momentum photon. The main idea of this Letter is to propose
a spontaneous breaking of translational symmetry to enable this process.

The resulting state has no currents and the variation of density $\delta n(\r)\ll \langle n\rangle$
is periodic in space $\delta n(\r+j_1\t_1+j_2t_2)=\delta n(\r)$ (here $\t_{1,2}$ are the primitive translation vectors)
It can therefore be classified as a {\em supersolid} state. If the primitive vectors of the reciprocal lattice $\b_{1,2}$
have the length of $k_F$ the excitation of the lowest state becomes allowed and the logarithmically divergent
negative correction to the ground state is present. We will see that this correction can overcome the positive contribution to the ground state energy from the compressibility
$(1/2)(\delta n)^2(\partial^2 E_{GS}/\partial n^2)_{F=0}$ thus making the supersolid state energetically favorable.

{\em Supersolid state and the phase diagram}--
In the presence of the periodic density variation, the fermionic gap in \req{spectrum-expanded} also acquires a spatial variation,
$\tilde \Delta \equiv (\partial \Delta/\partial n)_{n=\langle n\rangle} \delta n$. The correction \rref{correction1}  changes
due to the effect of the periodic potential produced by the supersolid:
\begin{subequations}
\be
\delta^{(n)}E_{GS}=-\sum\limits_{\b,\alpha}\ \int\limits_{\k \in BZ} \!\!\frac{d^2 k }{(2\pi)^2}\cE_{\b,\alpha}(\k);
%\frac{F_\sigma(\k,\b,j) F^*_\sigma(\k,\b,j)}{(\k+\b)^2/(2m)+\xi_j(\k)},
\ 
\cE=-
\frac{ |F_\sigma^{\b,\alpha}(\k)|^2}{\xi_{\b,\alpha}(k)},
\label{correction2}
\ee
where the quasimomentum integration is performed within the first Brillouin zone, $\b$  is a vector of the reciprocal lattice,
$\alpha$ labels the band for the $\gamma$  fermion in the periodic potential, described by Schr\"odinger equation
for the Bloch functions, $u_{j,\k}(\r+\t)=u_{j,\k}(\r)e^{i\k\t}$, 
\be
\left[{\beta\left(k_F^2+\nnabla^2\right)^2}/{4}+ \tilde{\Delta}(\r)\right] u_{\alpha,\k}(\r)
=\tilde{\xi}_\alpha (\k) u_{\alpha,\k}(\r).
\label{Bloch}
\ee
The energy of the two particle virtual state is 
$\xi_{\b,\alpha}(\k)\equiv d-\Lambda + (\k+\b)^2/(2m) +\tilde \xi_\alpha(\k)$, and the matrix elements
connecting excited states to the ground state are
\be
F_\sigma^{\b,\alpha}(\k)\equiv F_\sigma/{S_{uc}} \int_{uc}{d^2r}e^{-i(\k+\b)\cdot\r}u_{\alpha,\k}(\r).
\label{matrix-elements}
\ee
The integration is within the lattice unit cell of area $S_{uc}$, and Bloch functions
are normalized as
\be
\int_{uc} d^2 r|u_{\alpha,\k}(\r)|^2=S_{uc}.
\ee
\label{bandstructure}
\end{subequations}

For $\tilde\Delta=0$, equations \rref{bandstructure} are nothing but the expression \rref{correction1} folded into the first
Brillouin zone, as all the other couplings  \rref{matrix-elements} vanish. 

\begin{figure}[h]
\includegraphics[width = \columnwidth]{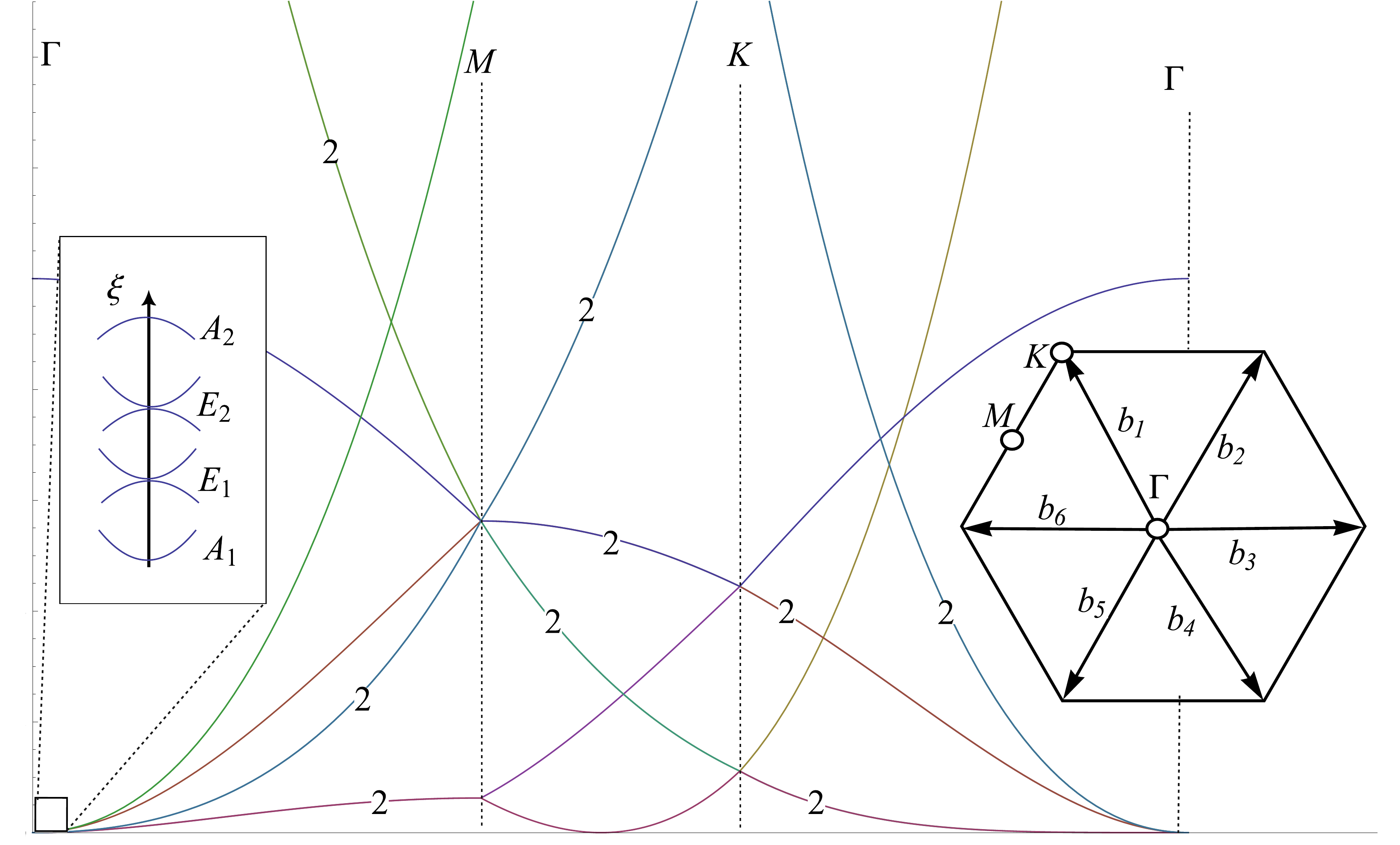}
\caption[The weak coupling spectrum of the $\gamma$-fermions in the periodic potential]{The weak coupling spectrum of the $\gamma$-fermions in the periodic potential.
The zoom shows the band structure near $\Gamma$ point. The labeling corresponds to the irreducible
representations of the point symmetry group ${\cal D}_6$. Right inset: The basis vectors for the reciprocal lattice $\b_{1,\dots, 6}$ and the first Brillouin zone.}
\label{fig2}
\end{figure}

For small 
$\tilde \Delta$ the relevant part of the spectrum can be described in the almost free particle approximation. Consider triangular lattice
\be
\tilde{\Delta}(r) =-\tilde{\Delta} \sum_{l=1}^{6}\exp(i\b_l\cdot \r),
\label{lattice-potential}
\ee
where the vectors $b_l$ are shown on Fig.~\ref{fig2} a).

On symmetry grounds only $A_1$ state,
 invariant under the symmetry group (see Fig.~\ref{fig2}), can contribute to the matrix elements \rref{matrix-elements} and
\be
F_\sigma^{0,A_1}(\k=0)=-\sqrt{6} F_\sigma\tilde{\Delta}/\Lambda,\quad \tilde{\xi}_{A_1}(0)=
-2\tilde{\Delta}.
\label{weak-coupling}
\ee
The linear in $\tilde{\Delta}$ shift of the lowest energy level $\tilde{\xi}_{A_1}$  is the signature of the triangular symmetry, ${\cal D}_6$;
the shift makes this lattice the most energetically profitable in comparison with, {\em e.g.}, square one.

The main contribution to the energy differences because the symmetry broken and the symmetric states
comes from the lowest energy part of the spectrum.
For the calculation with logarithmic accuracy,
the partial contribution $\cE$ can be approximately written as $\cE\approx d-\Lambda -2\tilde{\Delta}+k^2/(2m)$.
It yields
\be
\left[\delta^{(n)}\!-\delta^{(0)}\!\right]\! E_{GS}\!
= \frac{3\tilde{\Delta}^2m}{\pi}\!
\left[-\frac{|F_\sigma^2|}{\Lambda^2}Y(\tilde{d}-2\tilde{\Delta})+g
\right],
\label{energyfunction}
\ee
where   the detuning from the lowest excitation energy is given by
$\tilde{d}\equiv d-\Lambda$, and function $Y$ is defined as
\be
Y(X)= \ln\left[{k_F^2}/({m X})\right].
\label{yx}
\ee
The last term in \req{energyfunction} is the 
compressibility contribution and the positive constant, 
\be
g\equiv ({\pi}/{m})\left({\partial^2 E_{GS}}/{\partial n^2}\right)_{F=0}
\left({\partial \Delta}/{\partial n}\right)^{-2},
\ee
is of the order of unity. 

The correction given in \req{energyfunction} is the main result for the ground state energy at the lowest order in $|F|^2$.
 It shows that the broken symmetry {\em supersolid} state is always  energetically profitable for any finite $F$. However the potential is apparently pathological, as $Y(X)$ diverges as $X\rightarrow0$.  This infinite growth is an artifact of the lowest in $F$ approximation, as the presence of the 
external field leads to the level repulsion.
This level repulsion cuts the logarithm, and we turn to study of 
this effect.

 For $|F_\sigma|\ll \lambda$ it is sufficient to take into account only the two lowest energy states which couple to the radiation and
their interaction with the reference state without fermions. Then, the partial energy $\cE$ from \req{correction2},
with  account of \req{weak-coupling}, becomes the lowest
eigenvalues of the three state effective Hamiltonian
\be
\hat{H}_{eff}\simeq \begin{pmatrix}
d+\frac{k^2}{2m} & 0 & F_\sigma\\
0& d-\Lambda-2\tilde{\Delta}+\frac{k^2}{2m} & - \frac{\sqrt{6} F_\sigma\tilde{\Delta}}{\Lambda} \\
 F_\sigma^* & - \frac{\sqrt{6} F_\sigma^*\tilde{\Delta}}{\Lambda} & 0
\end{pmatrix}.
\label{Heff}
\ee
Straightforward calculation leads 
to the replacement 
\be
Y(\tilde{d}-2\tilde{\Delta})\to Y\left[\sqrt{\left(\tilde{d}(F_\sigma)-2\tilde{\Delta}\right)^2
+\frac{6\tilde{\Delta}^2|F_\sigma|^2}{\Lambda^2}
}\,
\right]
\label{energyfunctionprime}
\tag{\ref{energyfunction}$^\prime$}
\ee
in \req{energyfunction}, where $\tilde{d}(F_\sigma) \equiv d-\Lambda+ |F_\sigma|^2/\Lambda$
has the meaning of the lowest energy of the two-fermionic excitations shifted by the RF field.
\begin{figure}
\includegraphics[width = 0.7\columnwidth]{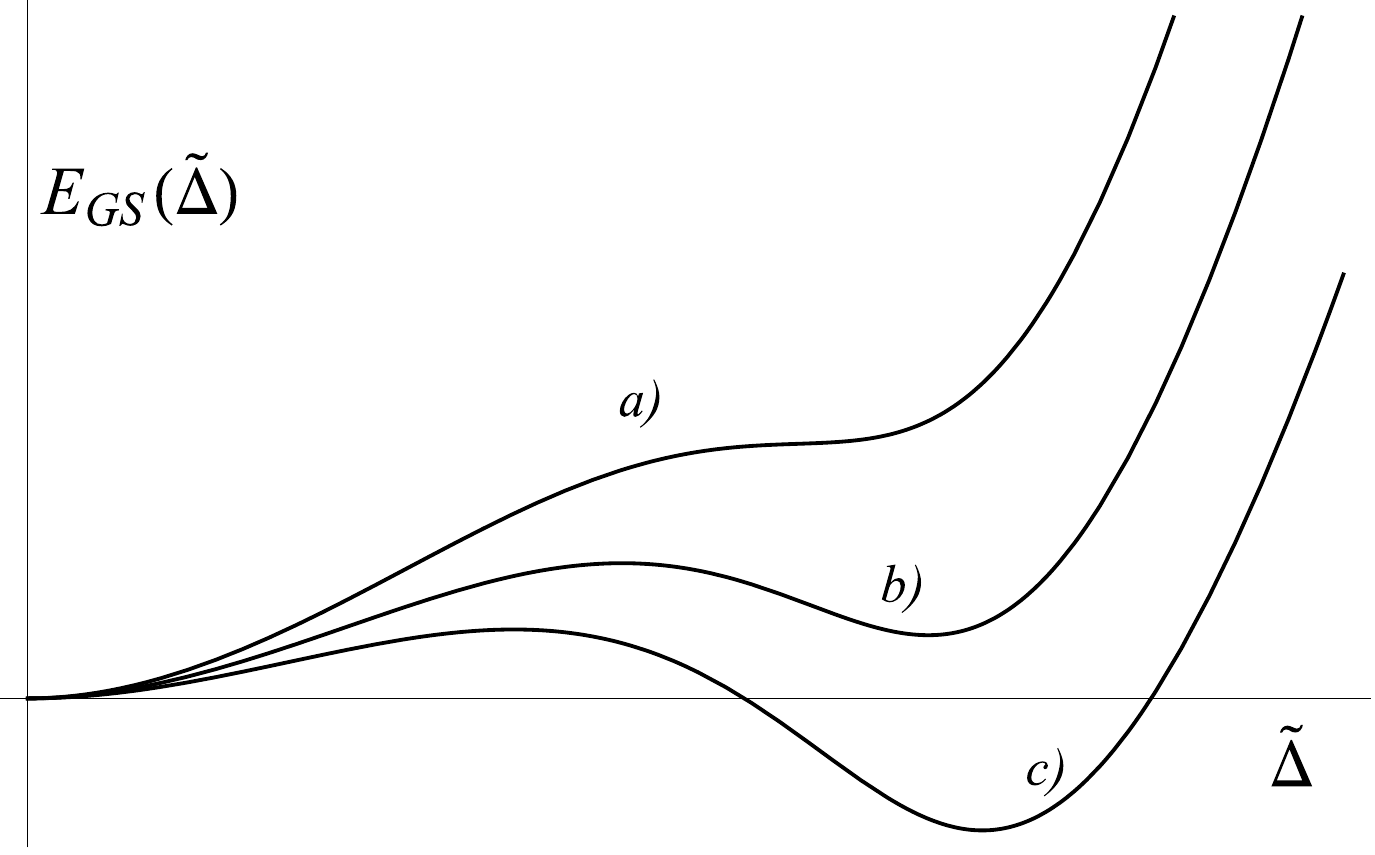}
\caption{
The correction to the ground state energy as a function of the order parameter, for liquid a) and supersolid c) states.
Though such energy profile is typical for the first order phase transitions, its functional form is
different from the Landau type expansion. 
}
\label{fig:energyprofile}
\end{figure}

\begin{figure}
\includegraphics[width = \columnwidth]{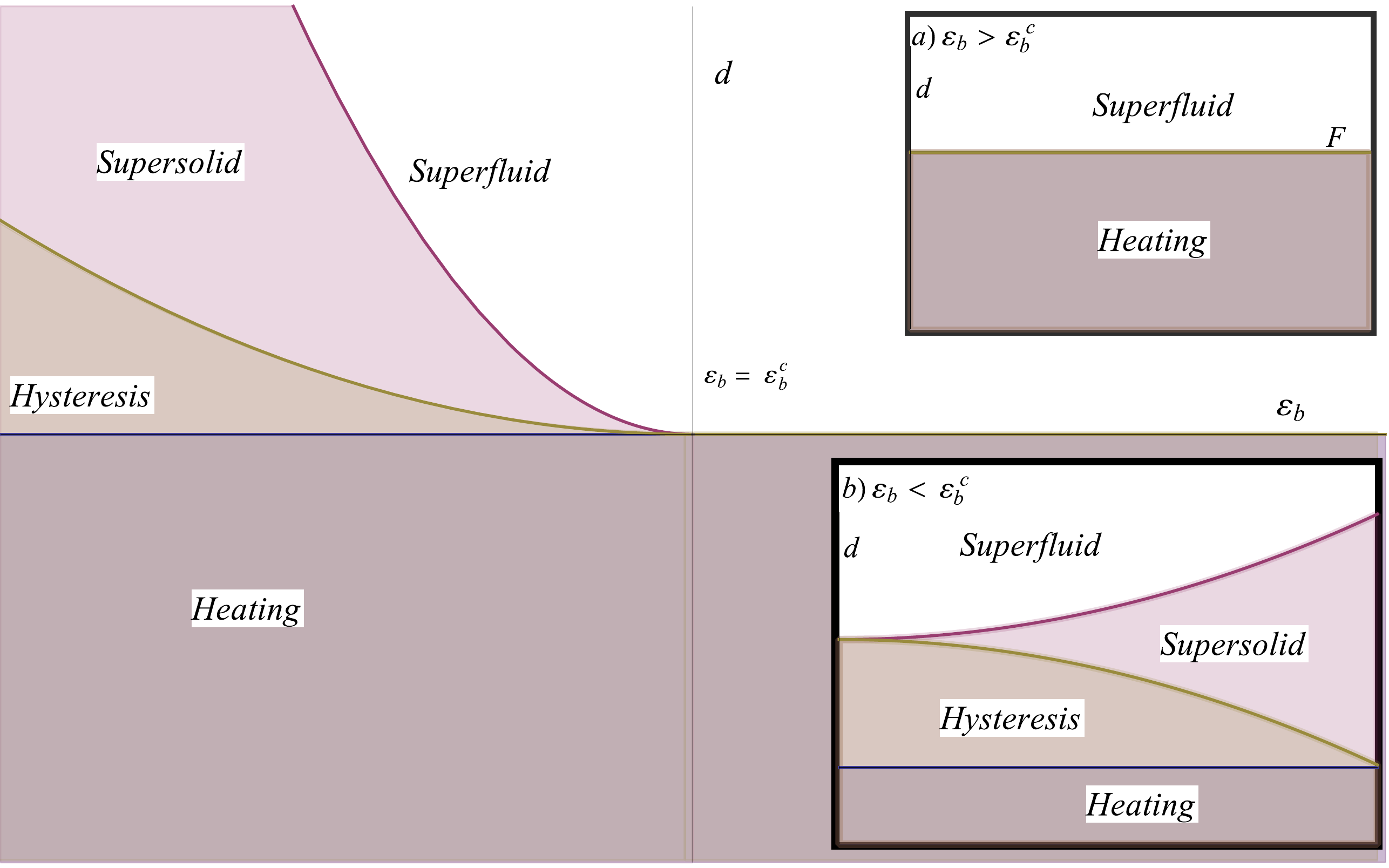}
\caption{
The proposed phase diagram at the fixed amplitude of the radiation field (main panel).
The insets show the phase diagram at fixed position of the 
resonance $\epsilon_b$ in the BEC a) and BCS b) regions.
The ``hysteresis'' region of the phase diagram depends
on the preparation of the system and may correspond either
to super-solid or the non-stationary normal phase heated by r.f.-radiation. 
}
\label{fig:phasediagram}
\end{figure}

The resulting form of the energy profile \rref{energyfunction}, \rref{energyfunctionprime} is shown
on Fig.~\ref{fig:energyprofile}. It shows two locally stable state characteristic of the {\em first order} phase transition.
Direct inspection shows that the supersolid state becomes more energetically profitable when 
$|\tilde{d}(F_\sigma)|\leq \tilde{d}_c(F_\sigma)$  where the critical detuning is given by
\be
Y\left(\frac{\sqrt{3}\tilde{d}_c(F_\sigma)|F_\sigma|^2}{2\Lambda^2}\right)
=\frac{\Lambda^2 g}{|F_\sigma|^2},
\label{dcritical}
\ee
The resulting phase diagram is shown on Fig.~\ref{fig:phasediagram}.

{\em In conclusion},
we have noted the BCS-BES crossover is necessarily followed by the reconstruction of topology of the spectrum of the fermionic excitations
and the critical field can be rigorously defined as the point of such change. We suggested an experimental scheme which transmutes this reconstruction of the 
{\em excitation} spectrum into a change in the symmetry of the {\em ground state}. For this scenario the {\em supersolid} state
is predicted to form.

The actual process by which the supersolid state forms is apparently quite complex. The process is an inherently non-equilibrium, zero temperature and first order phase transition. Each of these features alone bring interesting facets to the issue of the phase transition kinetics. Therefore this transition could be an interesting arena for testing theories of phase transition kinetics. 

We are grateful to O. Agam and L.I. Glazman for reading the
manuscript and to A.D. Vlasov for participation in the initial stage of 
this  project.
This work was supported by Simons foundation.

\newpage

\section{
Supplemental material: 
Hydrodynamics in the narrow resonance limit.}

The purpose of this supplementary section is to obtain
explicitly express the parameters of the hydrodynamic description
[Eqs. (2) -- (3) of the main text].
\setcounter{equation}{0}
We restrict ourselves to the two-dimensional case, $D=2$,
in the narrow resonance regime,
\be
\eta \equiv \lambda \sqrt{n}\left(\frac{n}{m}\right)^{-1} \ll 1.
\slabel{S0}
\ee
Moreover, we will consider the position  of the resonance $\varepsilon_b$
near the critical one $\varepsilon_b^c$ as discussed in the main text.

We begin with the Hamiltonian density (1) of the main text, setting all the gauge fields to zero.
\begin{equation}
\begin{aligned}
\hat{H} &= b^\dagger 
\left(\frac{-\nabla^2}{4m}  -\varepsilon_b\right) b  
+c^{\dagger}_\sigma\left( \frac{-\nabla^2}{2m}\right)c_\sigma\\
&+\frac{\lambda}{2} \left[b c^\dagger_{\sigma_1} \tau^y_{\sigma_1\sigma_2}c^\dagger_{\sigma_2} +h.c.\right].
\end{aligned}
\slabel{S1}
\end{equation}
 
As we are dealing with two-dimensional systems, all
the observable quantities expressed via the bare parameters of the Hamiltonian
contain logarithmic divergences, however, the relations
between different observables are free of such divergences.
To illustrate this point,
we note that the quantity $\varepsilon_b$ is not the observable location of the resonance. 
We may calculate the physical resonance $E_r$, 
by computing the correction to the energy  of one $b$ particle at zero momentum
due to the excitations of two virtual fermions:
\begin{equation}
\begin{aligned}
\delta\Pi\left(\omega\right)  & = -\lambda^2\!\! \int\!\!\frac{d^2k}{(2\pi)^2}\frac{1}{\frac{k^2}{m} - \omega} = -\frac{m\lambda^2}{4\pi}\ln\left(\frac{\Lambda}{\omega}\right),
\end{aligned}
\slabel{S2}
\end{equation}
where $\Lambda$ is an unphysical high energy cutoff.

The physical location of the resonance is determined by the self-consistency equation
\begin{equation}
E_r = - \epsilon_b -\frac{m\lambda^2}{4\pi}\ln\left(\frac{\Lambda}{|E_r|}\right),
\slabel{S3}
\end{equation}
where $\Lambda$ is some high-energy cut-off.
The value of $E_r$ is an observable position of the bound state at $E_r< 0$ and
the position of the resonance at $E_r>0$.
 
We proceed to calculate the properties of the ground in the in the saddle point approximation, which is valid in the narrow resonance regime. We take the spatially homogeneous ansatz
\begin{equation} 
b = \Delta/ \lambda \in \mathbb{R}.
\slabel{S4}
\end{equation}
and introduce the thermodynamic potential density so that
\be
E_{GS}(n)=\Omega(\mu)-\mu\partial_\mu\Omega(\mu),\slabel{S40}
\ee
where $\mu(n)$ is found from
\be
n =-\partial_\mu\Omega(\mu).
\slabel{S41}
\ee

The the thermodynamic potential
$\Omega(\mu)$ is found as
the ground state of the mean-field version of the Hamiltonian \rref{S1} 
\be\begin{aligned}
\hat{H} &= -\left(2\mu+\varepsilon_b\right)\frac{\Delta^2}{\lambda^2}  
+c^{\dagger}_\sigma\left( \frac{-\nabla^2}{2m}-\mu\right)c_\sigma\\
&+\left[\frac{\Delta}{2}c^\dagger_{\sigma_1} 
\tau^y_{\sigma_1\sigma_2}c^\dagger_{\sigma_2} +h.c.\right].
\end{aligned}
\slabel{S42}
\ee

After Bogoliubov rotation of the fermion operators in \req{S42}, we have their spectrum  for small $k$ and small $\mu$ is
\begin{equation}
\begin{aligned}
\epsilon(k) & = \left[ \left(\frac{k^2}{2m} - \mu\right)^2 + \Delta^2\right]^{1/2}\\
&\approx |\Delta| - \frac{k^2 \mu}{2m|\Delta|} +\frac{k^4}{8m^2|\Delta|}.
\end{aligned} 
\slabel{S5}
\end{equation} 
Comparing \rref{S5} with Eq. (3b) of the main text
we see that the critical point $\epsilon_b^c$ is determined by $\mu = 0$. 
Since we are investigating the vicinity of the region around this critical point we can restrict ourselves small $\mu$ limit.

The thermodynamic potential at zero temperature is given by,
\begin{equation}
\begin{aligned}
\Omega(\mu,\Delta)
 &=\left(-\varepsilon_b - 2\mu\right)\frac{\Delta^2}{\lambda^2}\\ &+ \int\frac{d^2 k}{(2\pi)^2}\left\{ \frac{k^2}{2m} - \mu - \left[\left(\frac{k^2}{2m} - \mu\right)^2 +\Delta^2\right]^{\frac{1}{2}}\right\}\\
 & \approx\left(-\varepsilon_b - 2\mu\right)\frac{\Delta^2}{\lambda^2} 
- \frac{m}{4\pi} \Delta^2\ln\left(\frac{\sqrt{e}\Lambda}{\Delta}\right)
+{\cal O}(\mu\Delta)
\\
& =\left(E_r - 2\mu\right)\frac{\Delta^2}{\lambda^2} -
 \frac{m}{4\pi} \Delta^2\ln\left(\frac{\sqrt{e}|E_r|}{\Delta}\right)+{\cal O}(\mu\Delta),
\end{aligned}
\slabel{S6}
\end{equation}
where in the last line the physical resonance $E_r$  from \req{S3} is used
to obtain the expression free of the logarithmic divergences. 

Minimizing \req{S6} with respect to $\Delta$ gives, 
\[
E_r - 2\mu = \frac{m\lambda^2}{4\pi}\ln\left(\frac{|E_r|}{\Delta}\right),
\]
with the resulting expression for the fermionic gap
\be
\Delta(\mu)=\Delta(0)\exp\left(\frac{8\pi\mu}{m\lambda^2}\right); 
\
 \Delta(0)= |E_r| \exp\left(-\frac{4\pi E_r}{m\lambda^2}\right)
\slabel{S70}
\ee
and the thermodynamic potential
\be
\Omega(\mu)=-\frac{m}{8\pi} \Delta^2(\mu);
\slabel{s71}
\ee

Varying $\mu$ and enforcing \req{S41}, we obtain
\begin{equation}
n = \frac{2\Delta^2(\mu)}{\lambda^2}
\slabel{S8}
\end{equation}
and the equation of states \rref{S40} as
\be
E_{GS}(n)=\frac{mn\lambda^2}{16\pi}\ln\left(\frac{n\lambda^2}{2{e}\Delta_0^2}\right);
\slabel{S81}
\ee

As noted above the critical detuning $\epsilon_b^c(n)=-E_r$ is defined by $\mu = 0$. 
Equations \rref{S70} and \rref{S8} give 
\[
\epsilon_b^c(n) = -\frac{m\lambda^2}{4\pi}\ln\left(\frac{\sqrt{2}\epsilon_b^c(n)}
{\lambda\sqrt{n}}\right)
%\frac{m\lambda^2}{4\pi}\log\left(\frac{2\epsilon_b^c_c(n)}{\lambda\sqrt{n}}\right).
\]
which with the logarithmic accuracy yields.
\be
\epsilon_b^c(n)=\frac{m\lambda^2}{4\pi}\ln\left(\frac{2^{3/2}\pi}{\eta}\right)
\slabel{S9}
\ee

The values of the remaining parameters entering into 
Eq. (3b), thus are [see \req{S5}]
\begin{equation}
\Delta=\frac{\lambda \sqrt{n}}{\sqrt{2}},\
\alpha =  \frac{1}{4m\Delta},\
\beta = \frac{1}{8m^2\Delta}.
\slabel{S10}
\end{equation}

\end{document}